\documentclass[11pt, a4paper]{article}

\usepackage{geometry}
\usepackage{setspace}
\usepackage[T1]{fontenc}
\usepackage{textcomp}
\usepackage{XCharter}
\usepackage[cmintegrals, cmbraces]{newtxmath}
\usepackage{FiraSans}
\usepackage[scale=0.85]{FiraMono}
\usepackage{microtype}

\usepackage[dvipsnames, svgnames, table]{xcolor} 
\definecolor{primaryblue}{HTML}{1A5276}
\definecolor{secondaryblue}{HTML}{5DADE2}
\definecolor{citegreen}{HTML}{0E6655}
\definecolor{maintext}{gray}{0.15}
\definecolor{metagray}{gray}{0.5}
\definecolor{boxbg}{gray}{0.98}

\color{maintext}

\usepackage{titlesec}
\titleformat{\section}
  {\normalfont\sffamily\bfseries\Large\color{primaryblue}}
  {\thesection}
  {0.8em}
  {}
  [{\vspace{2pt}\color{primaryblue!25}\titlerule[0.8pt]\vspace{10pt}}]
\titleformat{\subsection}
  {\normalfont\sffamily\bfseries\large\color{primaryblue}}
  {\thesubsection}
  {0.8em}
  {}
\titleformat{\subsubsection}
  {\normalfont\sffamily\bfseries\color{primaryblue}}
  {\thesubsubsection}
  {0.8em}
  {}
\titlespacing*{\section}{0pt}{24pt}{12pt}
\titlespacing*{\subsection}{0pt}{20pt}{10pt}

\usepackage{hyperref}
\hypersetup{
    colorlinks=true,
    linkcolor=primaryblue,
    citecolor=citegreen,
    urlcolor=primaryblue,
    pdftitle={Fine-Tuning Vision-Language Models for Markdown Conversion of Financial Tables in Malaysian Audited Financial Reports},
    pdfauthor={Jin Khye Tan, En Jun Choong, Ethan Jeremiah Chitty, Yan Pheng Choo, John Hsin Yang Wong, Chern Eu Cheah},
    pdfsubject={Automated Table Extraction from Financial Reports},
    pdfkeywords={VLM, Fine-Tuning, Table Structure Recognition, Markdown Conversion, Document AI, Deep Learning, Financial Reports, Malaysian Financial Reports},
    breaklinks=true,
    pdftoolbar=true,
    pdfpagemode=UseOutlines,
    unicode=true,
    pdfstartview={FitH},
}

\usepackage{graphicx}
\graphicspath{{media/}}
\usepackage{booktabs}
\usepackage{multirow}
\usepackage{amsmath}
\usepackage[labelfont={bf,sf,color=primaryblue}, textfont={sf,it}, font=small]{caption}
\usepackage{subcaption}
\usepackage[most]{tcolorbox}
\usepackage{fancyhdr}
\usepackage{natbib}
\usepackage{enumitem}
\setlist[itemize]{label=\textcolor{primaryblue}{\textbullet}, itemsep=2pt, topsep=4pt}
\setlist[enumerate]{label=\arabic*., itemsep=2pt, topsep=4pt}
\usepackage{float}
\usepackage{tabularx}
\usepackage{longtable}
\usepackage{ragged2e}
\usepackage{adjustbox}
\usepackage{fontawesome5}

\usepackage{listings}
\newtcolorbox{abstractbox}{
    enhanced,
    colback=boxbg,
    colframe=primaryblue!70!white,
    boxrule=0.8pt,
    fonttitle=\bfseries\sffamily\large,
    title=Abstract,
    sharp corners,
    pad at break=2mm,
    breakable
}

\begin{document}


    
    
        
        
        
    
    
    
    

\title{\sffamily\bfseries\color{primaryblue}\fontsize{20pt}{24pt}\selectfont Fine-Tuning Vision-Language Models for Markdown Conversion of Financial Tables in Malaysian Audited Financial Reports}

\author{
    \large\sffamily
    Jin Khye Tan\textsuperscript{1},
    En Jun Choong,
    Ethan Jeremiah Chitty,
    Yan Pheng Choo,
    John Hsin Yang Wong, and
    Chern Eu Cheah.
    \\[3ex]
    \small\sffamily
    \textsuperscript{1} Faculty of Computer Science and Information Technology, Universiti Malaya, 50603 Kuala Lumpur, Malaysia.
}

\makeatletter
\begin{center}
    {\color{primaryblue!25} \rule{\linewidth}{2pt}}
    \parbox{0.9\textwidth}{
        \centering
        \vspace{1em}
        \@title
        \vspace{0.5em}
    }
    {\color{primaryblue!25} \rule{\linewidth}{2pt}}

\end{center}

\begin{center}
    \parbox{0.9\textwidth}{
        \centering
        \@author
    }
\end{center}
\makeatother

\begin{abstractbox}
\noindent Accurately extracting and representing the structure of tabular data from financial documents remains a critical challenge in document understanding, particularly for regulatory and analytical use cases. This study addresses the complexity of converting financial tables from Malaysian audited financial reports into Markdown format, a task complicated by rotated layouts, multi-level headers, and implicit structural cues. We propose a fine-tuned vision-language model (VLM), based on Qwen2.5-VL-7B, optimized for high-fidelity Markdown generation from document images. Our approach includes a curated dataset of 2,152 image-text pairs with augmentations and a supervised fine-tuning strategy using LoRA. To assess performance, we evaluated our model on 100 out-of-sample tables using a dual framework: a criteria-based LLM-as-a-judge for fine-grained accuracy and our novel Markdown Tree-Edit-Distance-based Similarity (TEDS) metric for holistic structural fidelity. Our model achieves a \textbf{92.20\% overall accuracy}  on the criteria-based assessment and a \textbf{96.53\% Markdown TEDS score}. This performance significantly surpasses its Qwen2.5-VL-7B base model, larger-scale VLMs, and specialized reasoning-enabled models. Compared to these self-hosted alternatives, it also significantly reduces inference time. Furthermore, its accuracy exceeds that of widely used proprietary models such as OpenAI's GPT-4o and Gemini 2.5 Flash. These results demonstrate that domain-specific fine-tuning provides an effective and efficient method to bridge the gap between unstructured financial documents and downstream automation, rivalling much larger and more general models without their computational overhead.

\vspace{1em}


\end{abstractbox}


\clearpage

\section{Introduction}

\subsection{Background and Motivation}
Accurate extraction and structural representation of tabular data from unstructured documents remain central challenges in document understanding, commonly referred to as table structure recognition \citep{Schreiber2017DeepDeSRT, Siddiqui2019DeepTabStR}. Early approaches relied on rule-based systems and OCR engines such as Tesseract \citep{Smith2007Overview}, which often struggled with layout distortions and formatting variability. Subsequent advances introduced neural models such as TableNet \citep{Paliwal2019Tablenet} and GraphTSR \citep{Chi2019Complicated}, leveraging visual and spatial cues to infer table structures more robustly. 

More recently, vision-language models (VLMs) have emerged as powerful tools for document intelligence. Models such as LayoutLM \citep{Xu2020LayoutLM}, Donut \citep{Kim2022OCR}, and proprietary models, such as GPT-4o \citep{OpenAI2024GPT} and Gemini 2.5 \citep{Comanici2025Gemini}, jointly model textual, visual, and spatial information, enabling generalizable and high-fidelity table understanding.

\subsection{Challenges in Financial Table Understanding}
In the financial domain, accurate extraction of tabular data from reports is essential for tasks such as auditing, corporate analysis, and regulatory compliance. Financial reports typically contain structured tables in key sections such as income statements, balance sheets, cash flow statements, statements of changes in equity, and notes to the financial statements. These tables are central to assessing a company’s financial health and performance \citep{Rejison2025Identifying}. Preserving the structural integrity of these tables, including the correct alignment of headers, values, and contextual labels, is critical for enabling downstream applications such as ratio analysis, risk modeling, and automated reporting. As highlighted by \citet{Samantapudi2025Table}, even small structural errors can lead to significant issues in financial interpretation or regulatory compliance. High-accuracy automation of this process is therefore a fundamental requirement for scalable and reliable financial analysis, supporting use cases such as earnings evaluation, reconciliation, peer benchmarking, and market surveillance.

However, financial tables in real-world documents often vary significantly in layout, structure, and presentation. This variability is also observed in Malaysian audited financial reports, which commonly contain rotated tables, multi-level headers, implicit columns, and missing grid lines. These complexities pose significant challenges for document parsing, even for state-of-the-art VLMs. Section \ref{sec:challenges} provides illustrative examples demonstrating their impact on structured output generation.

\subsection{Our Approach: VLM Fine-Tuning for Malaysian Financial Reporting Adaptation}
To address these challenges, we introduce a "Markdownification" pipeline that converts financial tables into Markdown representations, guided by domain-specific formatting rules tailored to Malaysian financial reporting. This process flattens hierarchical headers, preserves multi-entity and multi-period distinctions, and explicitly identifies implicit structural elements (e.g., note indicators). By standardizing diverse layouts into a consistent textual representation, "Markdownification" enhances the reliability of downstream LLM-based financial analysis.

Initial experiments with open-source VLMs indicated suboptimal performance in interpreting complex financial tables, revealing significant structural errors. Proprietary models (Gemini 2.5, GPT-4o) showed improved results, yet still required extensive prompt adjustments and produced inconsistencies. Their closed-source nature and elevated costs further limit scalability and transparency, highlighting the need for customized, self-hosted solutions with strong domain expertise.

We present a fine-tuned adaptation of Qwen2.5-VL-7B-Instruct \citep{Bai2025Qwen2} optimized for Markdown generation from Malaysian financial tables. Our contributions are:

\begin{itemize}
        \item \textbf{Domain-specific Dataset}: A development set of 2,152 text-image pairs from financial statements and notes, including 30\% rotated augmentations, was used for training and validation. Final performance was measured on a held-out, distinct test set of 100 tables.
    \item \textbf{Fine-Tuning Methodology}: Supervised Fine-Tuning (SFT) with LoRA \citep{Hu2021LoRA} using the LLaMA-Factory framework \citep{Zheng2024LlamaFactory} on two units of A100 40GB GPUs.
    \item \textbf{Significant Accuracy Uplift}: Achieved a \textbf{92.20\% overall accuracy} on a criteria-based LLM evaluation and a \textbf{96.53\% Markdown TEDS score} for holistic structural fidelity. This performance dramatically surpasses not only the Qwen2.5-VL-7B base model (32.80\% accuracy, 52.08\% TEDS), but also larger models such as Qwen2.5-VL-32B (68.60\% accuracy, 71.20\% TEDS) and Qwen2.5-VL-72B AWQ (79.80\% accuracy, 74.72\% TEDS), reasoning-intensive models such as MiMo-VL-7B-RL (58.20\% accuracy, 58.56\% TEDS), and proprietary models such as Gemini 2.5 Flash (82.40\% accuracy, 79.19\% TEDS) and GPT-4o (65.20\% accuracy, 74.41\% TEDS).
\end{itemize}

\clearpage

\section{Challenges in Malaysian Financial Tables}
\label{sec:challenges}
Malaysian audited financial reports exhibit significant heterogeneity in table design, posing unique challenges for automated extraction. Even advanced VLMs struggle with accurate Markdown conversion due to layout irregularities. Below, we detail six key challenges, their implications, and mitigation strategies.

\begin{enumerate}
    \item \textbf{Inconsistent Table Formats}: Tables vary widely in structure, including column count, header depth, use of grid lines, even within the same document. This lack of standardization prevents the use of fixed parsing rules, requiring adaptive, context-aware models.
    
    \vspace{1em}
 
    \begin{minipage}{\linewidth}
    \centering
        \fbox{\includegraphics[height=4.5cm]{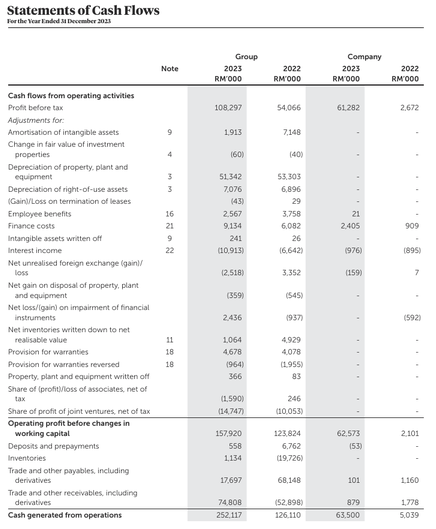}}
        \hspace{5em}
        \fbox{\includegraphics[height=4.5cm]{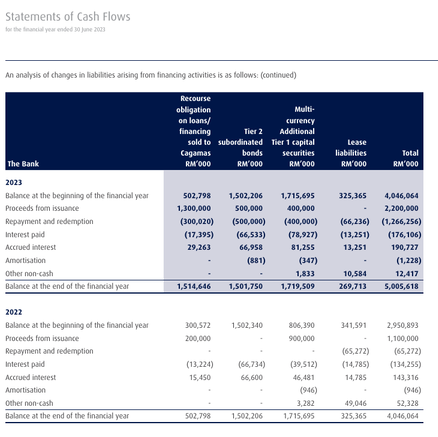}}
    \captionof{figure}{Variations in table formatting across different financial reports.}
    \label{fig:table_variations}
    \vspace{1em}
    \end{minipage}

    \item \textbf{Rotated Layouts}: Wide tables are frequently rotated 90\textdegree{} to fit page constraints. Such rotations disrupt a VLM's standard text-flow interpretation, causing models to misalign rows and columns.

    \begin{minipage}{\linewidth}
    \centering
    \fbox{\includegraphics[width=0.9\textwidth]{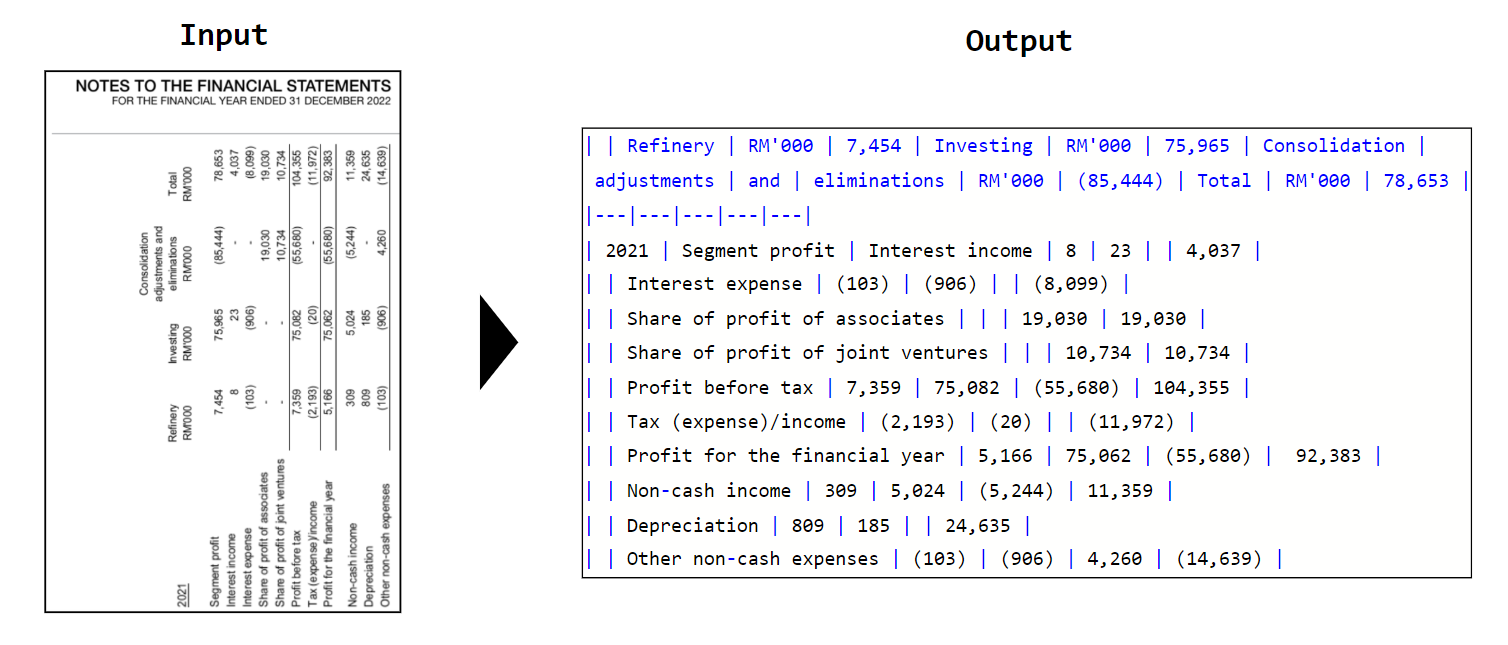}}
    \captionof{figure}{Example of a rotated table and its inaccurate Markdown output.}
    \label{fig:rotated}
    \vspace{1em}
    \end{minipage}

    Figure \ref{fig:rotated} shows how this orientation leads the model to misread the table structure, resulting in a transposed and inaccurate Markdown output.

    \clearpage

    \item \textbf{Multi-level Headers}: Hierarchical headers are common but incompatible with standard Markdown, which lacks native support for cells spanning multiple rows or columns. To preserve meaning, these headers must be flattened into descriptive single-line equivalents.
    
    \begin{minipage}{\linewidth}
    \centering
    \fbox{\includegraphics[width=0.9\textwidth]{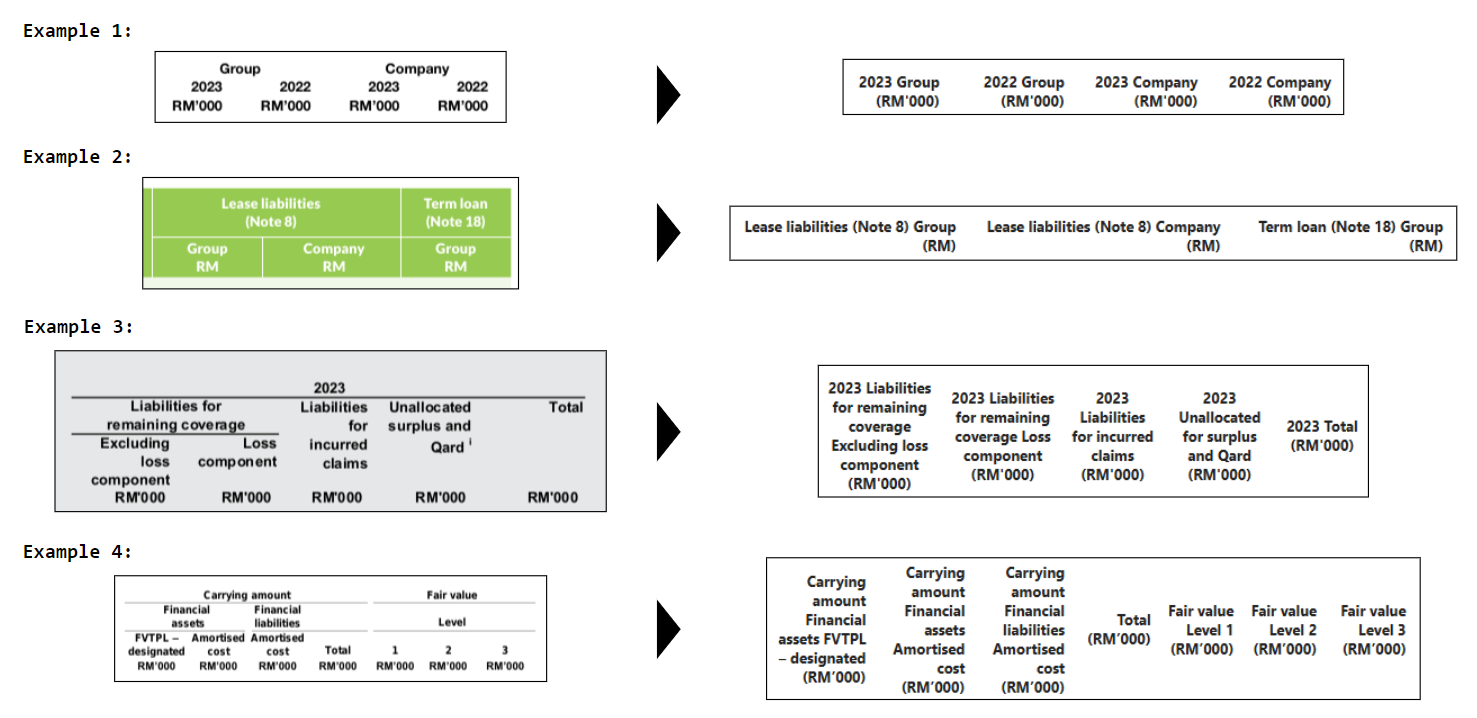}}
    \captionof{figure}{Example of flattening multi-level headers.}
    \label{fig:flatten_multilevel}
    \vspace{1em}
    \end{minipage}

    As illustrated in Figure \ref{fig:flatten_multilevel}, our solution 'flattens' these headers by concatenating the hierarchical information into a single descriptive title for each column, preserving the full semantic context.
    
    \item \textbf{Multi-Entity and Multi-Period Data}: Tables often contain data for multiple entities (e.g., "Group" and "Company") and periods ("2023", "2022"). Accurate parsing requires explicit association of each value with its correct entity and time period. Our Markdown schema ensures that every cell is unambiguously labeled, preventing misattribution in downstream analysis.

    \begin{minipage}{\linewidth}
    \centering
    \fbox{\includegraphics[width=0.9\textwidth]{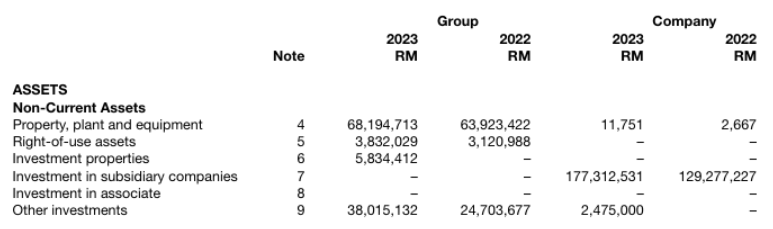}}
    \captionof{figure}{Table containing multi-entity and multi-period data.}
    \vspace{1em}
    \end{minipage}

    \clearpage
    
    \item \textbf{Lack of Grid Lines and Implicit Structures}: Many tables rely on whitespace and alignment rather than explicit borders. This ambiguity can cause VLMs to misalign values or misinterpret the structure.

    \begin{minipage}{\linewidth}
    \centering
    \fbox{\includegraphics[width=0.85\textwidth]{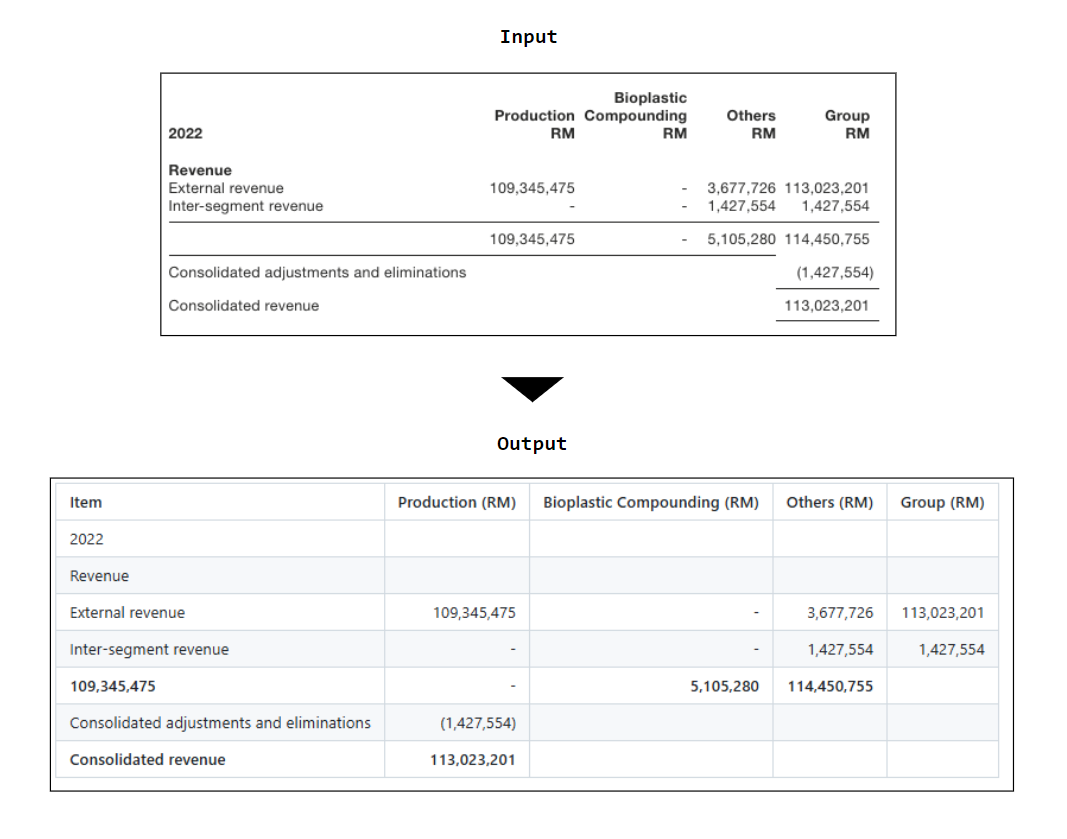}}
    \captionof{figure}{Examples illustrating structural ambiguity in tables without clear gridlines.}
    \label{fig:gridline_ambiguity}
    \vspace{1em}
    \end{minipage}

    In Figure \ref{fig:gridline_ambiguity}, the indented totals and implicit column breaks are clear to a human reader but confuse the VLM, leading to the inaccurate output shown.
    
    \item \textbf{Missing or Ambiguous Column Headers}: Some columns lack explicit headers. For example, note indicators such as "(a)", "(b)" function as a de facto "Note" column but are not labeled as such. Without structural cues, VLMs may misplace these labels, appending them to values or omitting them entirely. In our dataset, we annotate such cases explicitly, training the model to output them as a dedicated column.

    \begin{minipage}{\linewidth}
    \centering
    \fbox{\includegraphics[width=0.9\textwidth]{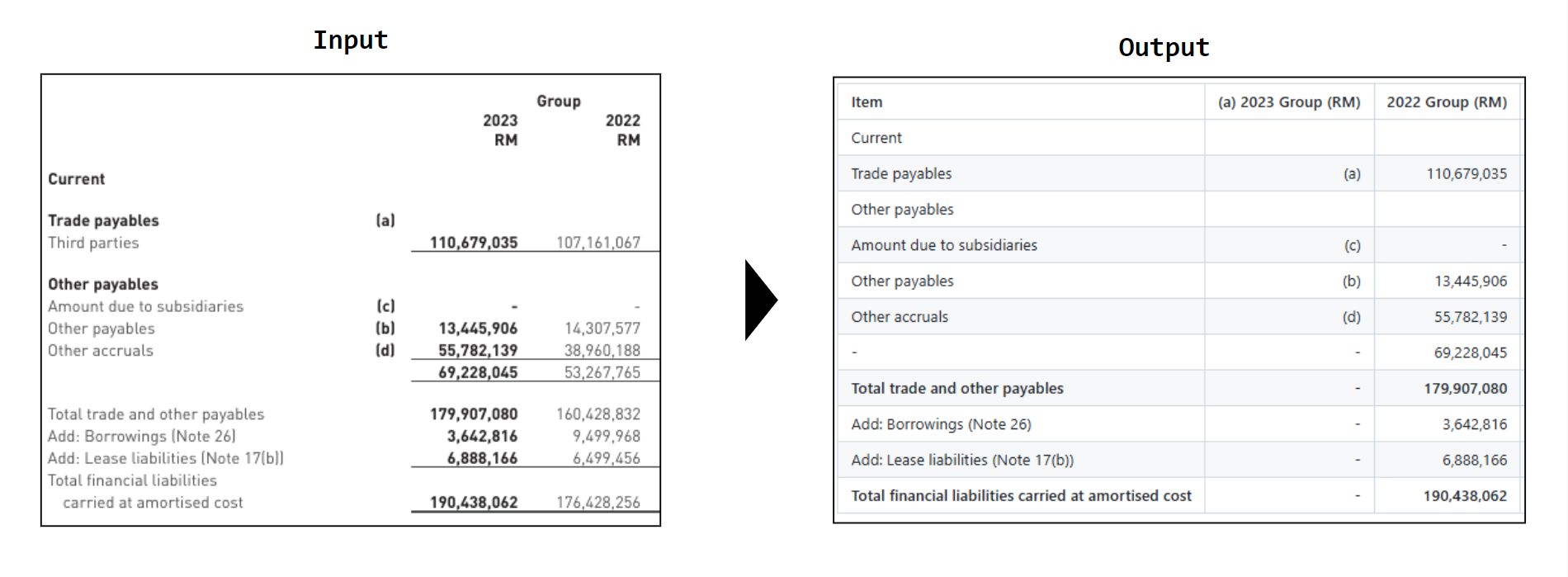}}
    \captionof{figure}{Example illustrating issues with missing or ambiguous column headers.}
    \vspace{1em}
    \end{minipage}
    
\end{enumerate}

\section{Related Works}
Recent advances in vision-language models (VLMs) have significantly improved document understanding, particularly in financial contexts \citep{Aida2025Enhancing}. By integrating layout, text, and visual features, modern VLMs surpass traditional OCR-based pipelines in robustness and accuracy. However, studies continue to show limitations in parsing complex financial tables, especially those with rotated layouts or implicit structures \citep{Aida2025Enhancing, Srivastava2025Enhancing}.

A growing body of work emphasizes the value of intermediate structured representations such as linearized tables or semantic markup for enhancing LLM reasoning on visual data \citep{Bradley2024SynFinTabs}. These formats act as bridges between raw document images and downstream analytical systems, improving fidelity in tasks such as financial data extraction and chart interpretation.

Nonetheless, VLMs still struggle to generate structured outputs such as Markdown that are reliable for downstream applications while also preserve hierarchical relationships and contextual semantics, which are critical for accurate financial analysis. These challenges are exacerbated by the heterogeneous and complex structure of tables in financial documents \citep{Balsiger2024Assessing}.

While models such as GPT-4o show promise in numeric and textual extraction, performance varies with document complexity. Recent research advocates for fine-tuning on domain-specific datasets to improve markup fidelity \citep{Bradley2024SynFinTabs}. Open-source frameworks such as olmOCR demonstrate that fine-tuned medium-sized VLMs can rival proprietary models in scalability and cost-efficiency \citep{Poznanski2025olmOCR}.

Our work aligns with and extends this direction by focusing on fine-tuning Markdown generation in the context of Malaysian financial reporting, where layout diversity and structural ambiguity are particularly acute.

\section{Methodology}
This section provides a comprehensive overview of our methodology for fine-tuning Qwen2.5 VL 7B for Markdown conversion, detailing the process from data acquisition to model evaluation. We outline the steps involved in constructing a specialized dataset, the strategy for data augmentation, the configuration for model training, and the metrics used for evaluating the model's performance in converting financial tables to Markdown.

\begin{figure}[H]
    \centering
    \fbox{\includegraphics[width=0.9\textwidth]{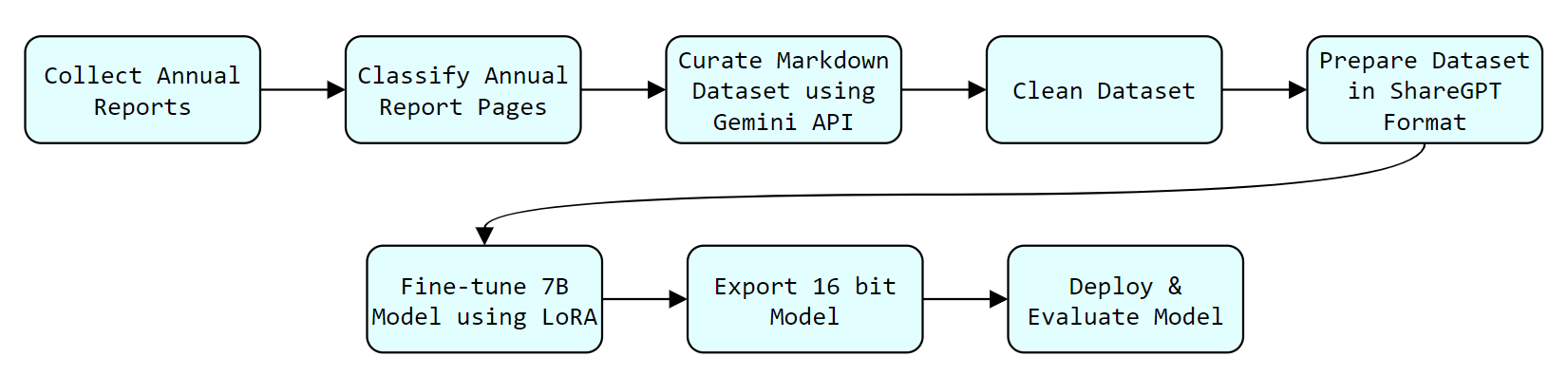}}
    \caption{Overview of our VLM fine-tuning pipeline.}
    \label{fig:pipeline_overview}
\end{figure}

The entire end-to-end workflow of our methodology is illustrated in Figure \ref{fig:pipeline_overview}. It outlines the major phases of the project, each of which is described in the following sections.

\subsection{Dataset Construction}

\begin{figure}[H]
    \centering
    \fbox{\includegraphics[width=0.9\textwidth]{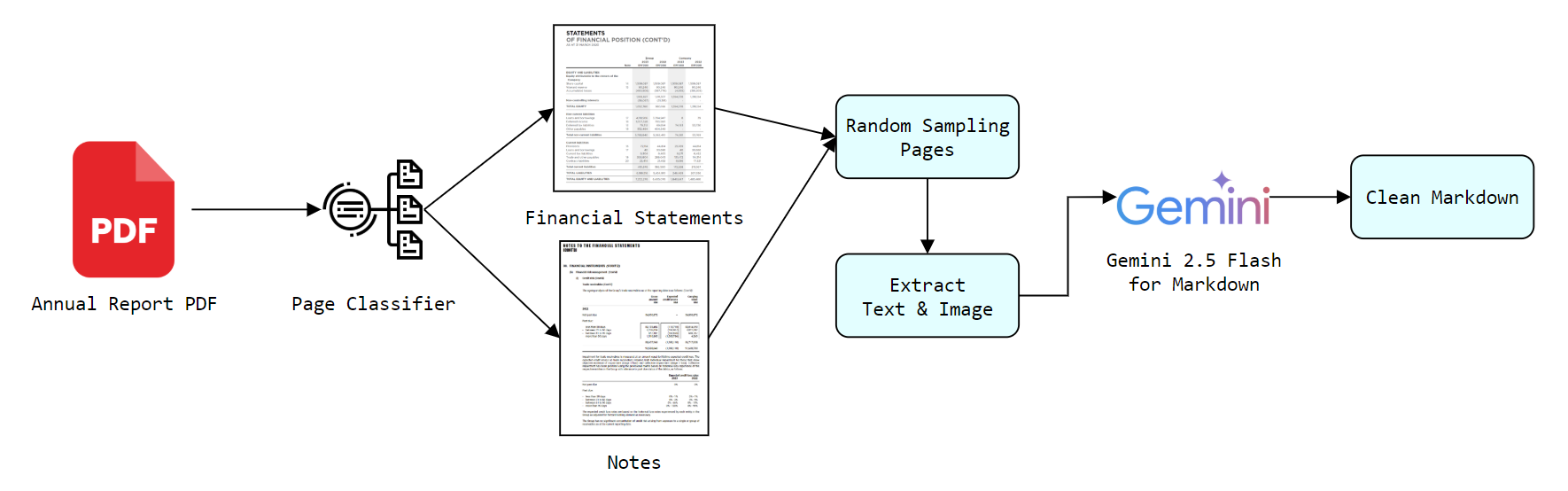}}
    \caption{Overview of the Markdown dataset creation process.}
    \label{fig:dataset_creation}
\end{figure}

The first phase of our work focuses on building the training dataset. Figure \ref{fig:dataset_creation} provides a visual summary of the key stages involved in this process, which are detailed below.

\subsubsection{Source Collection and Section Filtering}
We collected 991 Malaysian audited financial reports from public websites of Malaysian-listed companies, representing 741 unique entities across diverse sectors. To focus on content-rich financial data, we targeted two sections: Financial Statements and Notes to the Financial Statements. An XGBoost classifier~\citep{Chen2016-kv}, trained on TF-IDF~\citep{Salton1988-hd} features of page text, was used to automatically identify relevant pages. This approach leverages the robustness of gradient boosting models, which have been successfully applied to a range of natural language processing tasks, including personality trait classification from social media data~\citep{choong_predicting_2021}. Outputs were validated through random sampling and manual verification to ensure classification accuracy.

\subsubsection{Text and Image Extraction}
Our methodology for each selected page consists of the following steps:
\begin{itemize}
    \item Text was extracted using \texttt{pypdfium2}~\citep{pypdfium2-team}, which captures content but not layout fidelity, often resulting in misaligned labels and disordered blocks.
    \item Images were rendered at 100 DPI and encoded in base64 format to preserve visual structure.
\end{itemize}
These text-image pairs were processed by Gemini 2.5 Flash to generate initial Markdown outputs with our curated prompt (see Appendix~\ref{sec:appendix}). However, due to reliance on misaligned text extraction, outputs often reflected structural inaccuracies (e.g., transposed headers, missing columns).

\subsubsection{Manual Cleaning and Finalization}
All outputs were manually reviewed and corrected to ensure structural correctness. Key steps included:
\begin{itemize}
    \item Flattening multi-line headers into single descriptors.
    \item Correcting misaligned values and labels.
    \item Removing non-essential elements (footers, registration numbers).
    \item Standardizing Markdown formatting.
\end{itemize}
The cleaned dataset consisted of 1,656 high-quality samples as detailed in Table \ref{tab:dataset_initial}.

\begin{table}[H]
\centering
\caption{Initial Dataset Composition.}
\label{tab:dataset_initial}
\renewcommand{\arraystretch}{1.2}
\rowcolors{2}{boxbg}{white}
\begin{tabular}{@{}p{0.3\textwidth} p{0.45\textwidth} r@{}}
\toprule
\rowcolor{white} \textbf{Category} & \textbf{Details} & \textbf{Count} \\ 
\midrule
Financial Statements & Markdown outputs from main statement categories (Income, Balance Sheet, etc.). & 717 \\ 
Notes to the Financial Statements & Markdown outputs derived from the notes section of reports. & 742 \\
Hard Outlier Tables & Handpicked challenging tables to address edge cases. & 197 \\ 
\midrule
\rowcolor{white} \multicolumn{2}{l}{\textbf{Initial Dataset Total}} & \textbf{1,656} \\ 
\bottomrule
\end{tabular}
\end{table}

\subsection{Data Augmentation and Dataset Splits}
\label{sec:data_splits}

To improve the model's ability to handle rotated tables, a common layout in financial reports, we augmented the 1,656-sample dataset. This was done by creating rotated duplicates of 30\% of the existing entries (496 samples), which were randomly rotated by either 90\textdegree{} or 270\textdegree{}. This resulted in a final training and validation pool of 2,152 entries. Each entry in the dataset consists of three components:

\begin{enumerate}
    \item A 200 DPI image of the financial table page (either in its original orientation or rotated).
    \item The raw text extracted from the page using \texttt{pypdfium2}, which serves as part of the model's input prompt.
    \item The corresponding ground truth Markdown table, which serves as the target output for the model to learn.
\end{enumerate}
The entire dataset was structured in the ShareGPT format for LLaMA Factory framework compatibility (see Appendix~\ref{sec:appendix} for a format example).

The 2,152 samples constituted our development set, while a \textbf{held-out test set of 100 distinct tables} was curated for final performance evaluation. During the fine-tuning process in LLaMA Factory, a 10\% split of the 2,152-sample development set was used for validation to monitor for overfitting and guide model selection. The final performance of all models was exclusively evaluated on the 100-sample test set, which had zero overlap with any data seen during training or validation. The complete data partitioning is detailed in Table \ref{tab:dataset_splits}.

\begin{table}[H]
\centering
\caption{Dataset Partitioning for Training, Validation, and Testing.}
\label{tab:dataset_splits}
\renewcommand{\arraystretch}{1.2}
\rowcolors{2}{boxbg}{white}
\begin{tabular}{@{}p{0.3\textwidth} p{0.45\textwidth} r@{}}
\toprule
\rowcolor{white} \textbf{Set} & \textbf{Purpose} & \textbf{Count} \\ 
\midrule
Training Set & Used to fine-tune the model's weights. (90\% of the 2,152-sample development set) & 1,937 \\
Validation Set & Used during training to monitor performance and select the best checkpoint. (10\% of the 2,152-sample development set) & 215 \\
\rowcolor{boxbg} \textbf{Development Set Total} & & \textbf{2,152} \\
\midrule
Test Set & A completely distinct, held-out set used for final performance evaluation of all models. & 100 \\ 
\midrule
\rowcolor{white} \multicolumn{2}{l}{\textbf{Total Curated Samples}} & \textbf{2,252} \\ 
\bottomrule
\end{tabular}
\end{table}

\subsection{Training and Evaluation}
\subsubsection{Training Configuration}
The training process fine-tuned the Qwen2.5-VL-7B-Instruct model for Markdown generation from Malaysian financial tables. Based on token length analysis, we set \texttt{cutoff\_len = 6656} to ensure full sequences were accommodated without truncation. Due to GPU memory limitations, we used a LoRA (Low-Rank Adaptation) approach within the LLaMA-Factory framework. Training was conducted on two units of A100 40GB GPUs with an effective batch size of 16. All settings were chosen through trial and error during early experimentation.

\begin{table}[H]
\centering
\caption{Key Training Parameters for Fine-Tuning.}
\label{tab:training_params}
\renewcommand{\arraystretch}{0.9}
\begin{tabular}{@{}ll@{}}
\toprule
\textbf{Parameter} & \textbf{Value} \\ 
\midrule
\texttt{model\_name\_or\_path} & \texttt{Qwen/Qwen2.5-VL-7B-Instruct} \\
\texttt{template} & \texttt{qwen2\_vl} \\
\texttt{finetuning\_type} & \texttt{lora} \\
\texttt{cutoff\_len} & \texttt{6656} \\
\texttt{learning\_rate} & \texttt{5e-5} \\
\texttt{num\_train\_epochs} & \texttt{6} \\
\texttt{per\_device\_train\_batch\_size} & \texttt{1} \\
\texttt{gradient\_accumulation\_steps} & \texttt{8} \\
\texttt{lr\_scheduler\_type} & \texttt{cosine} \\
\texttt{warmup\_ratio} & \texttt{0.05} \\
\texttt{freeze\_vision\_tower} & \texttt{True} \\
\texttt{freeze\_multi\_modal\_projector} & \texttt{False} \\
\texttt{lora\_rank} & \texttt{64} \\
\texttt{lora\_alpha} & \texttt{64} \\
\texttt{lora\_dropout} & \texttt{0.1} \\
\texttt{val\_size} & \texttt{0.1} \\
\bottomrule
\end{tabular}
\end{table}

\begin{figure}[H]
    \centering
    \fbox{\includegraphics[width=0.9\textwidth]{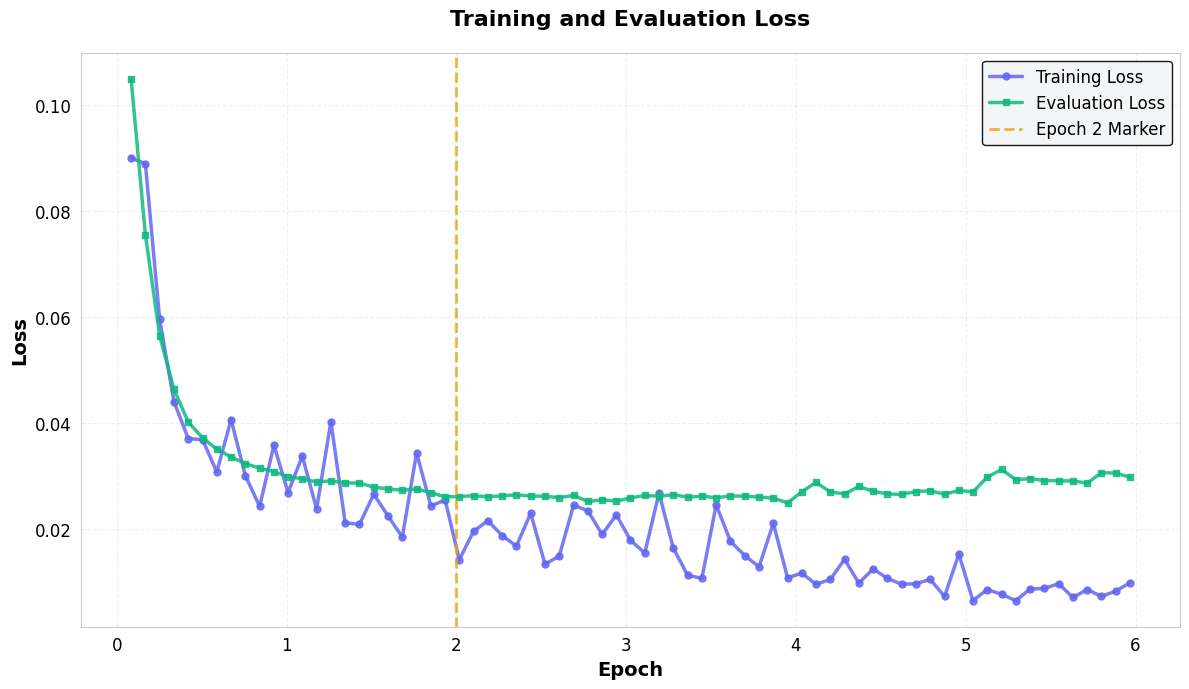}}
    \caption{Training and Evaluation Loss vs. Epochs.}
\end{figure}

Training ran for 6 epochs, with the optimal checkpoint selected at Epoch 2 based on manual evaluation of Markdown output quality, focusing on structural accuracy and semantic preservation. The model's learning trend over the six epochs is detailed in the training and evaluation loss graph. During the first epoch, a sharp, simultaneous drop in both training and evaluation loss indicates that the model was effectively learning the dataset's general patterns. After this initial phase, the curves diverge; the training loss continues its downward trend, while the evaluation loss flattens, reaching its minimum value around the second epoch before beginning to subtly increase in later epochs. This trend is characteristic of overfitting, where the model begins to memorize the training set at the expense of its ability to generalize. The observation that the evaluation loss was lowest at epoch 2 aligns with the decision to select this checkpoint to ensure optimal performance on unseen data.

\subsubsection{Evaluation Methodology}
\label{sec:eval_methodology}

To evaluate our fine-tuned Qwen2.5-VL-7B model, we conducted a comprehensive assessment using a test set of 100 financial tables extracted from Malaysian audited financial reports of 96 unique companies, comprising 50 tables from the Financial Statements section and 50 from the Notes to the Financial Statements section. The full benchmark dataset is available in Section~\ref{sec:data_availability}.

The evaluation process involved processing each table image through our "Markdownification" pipeline to generate raw Markdown outputs using the same curated prompt employed during dataset generation (see Appendix~\ref{sec:appendix}). These outputs were then compared against ground truth reference Markdowns, which were created by manually cleaning and refining the model-generated outputs to ensure structural and semantic accuracy.

Our evaluation framework is dual-faceted, leveraging two complementary approaches to capture both fine-grained correctness and overall document integrity.

\paragraph{Criteria-Based LLM-as-a-Judge}
First, we employed OpenAI's \texttt{o3-mini} as an automated judge to provide a structured, criteria-based comparison between the model's raw output and the ground truth. This LLM-as-a-judge method uses our structured prompt template (see Appendix~\ref{sec:appendix}), which defines five key evaluation criteria. To ensure clarity, Table \ref{tab:llm_judge_criteria} provides an example of an error that would cause a failure for each criterion:

\begin{longtable}{@{} p{0.5\textwidth} p{0.5\textwidth} @{}}

\caption{LLM Judge Evaluation Criteria with Failure Examples.}
\label{tab:llm_judge_criteria} \\

\toprule
\textbf{Criterion \& Description} & \textbf{Example of a Failure (Bad Output)} \\
\midrule
\endfirsthead

\multicolumn{2}{c}%
{{\bfseries \tablename\ \thetable{} -- continued from previous page}} \\
\toprule
\textbf{Criterion \& Description} & \textbf{Example of a Failure (Bad Output)} \\
\midrule
\endhead

\midrule
\multicolumn{2}{r@{}}{\textit{Continued on next page}} \\
\endfoot

\bottomrule
\endlastfoot


\RaggedRight
\textbf{Correct Row Count} \newline 
Output must have the same number of data rows as the ground truth. & 
\RaggedRight
\leavevmode\newline
The model omits several financial period rows and the "Group" row, resulting in an incomplete table. \\
\multicolumn{2}{c}{
    \includegraphics[width=\linewidth, keepaspectratio]{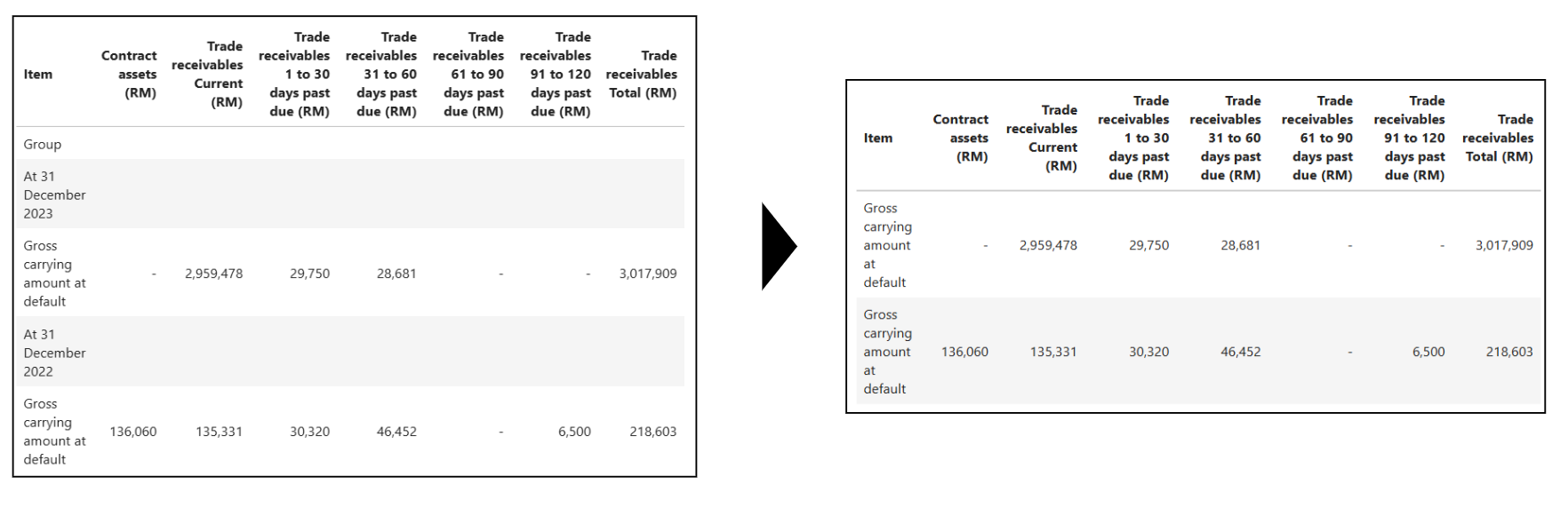}
} \\
\midrule

\RaggedRight
\textbf{Correct Column Count} \newline 
Output must have the same number of columns as the ground truth. &
\RaggedRight
\leavevmode\newline
The model fails to extract the "2023 Group" column, leading to a loss of an entire data series. \\
\multicolumn{2}{c}{
    \includegraphics[width=\linewidth, keepaspectratio]{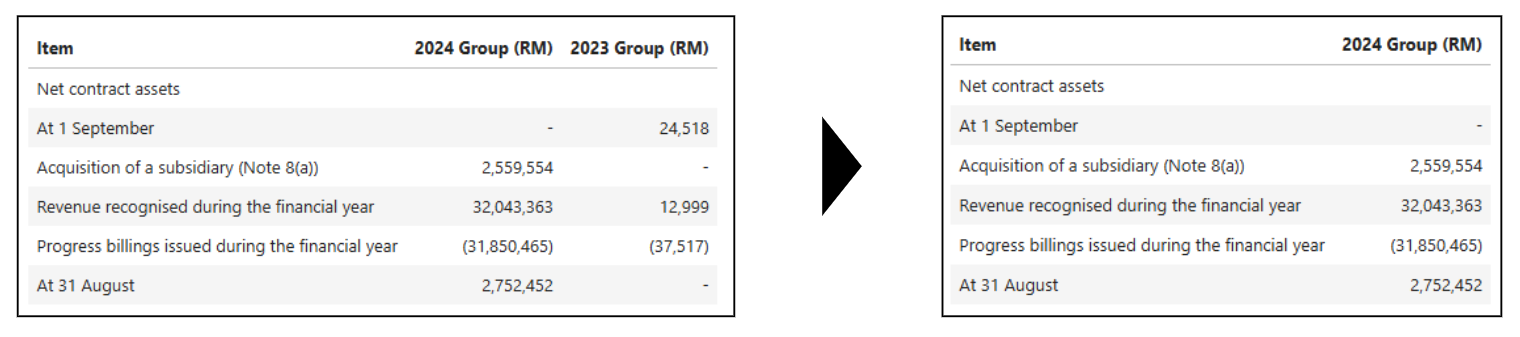}
} \\
\midrule

\RaggedRight
\textbf{Semantically Accurate Headers} \newline 
Header text must preserve the same meaning as the ground truth. &
\RaggedRight
\leavevmode\newline
The model incorrectly omits the "The Company" specifier from the headers, creating ambiguity about the data's entity level. \\
\multicolumn{2}{c}{
    \includegraphics[width=\linewidth, keepaspectratio]{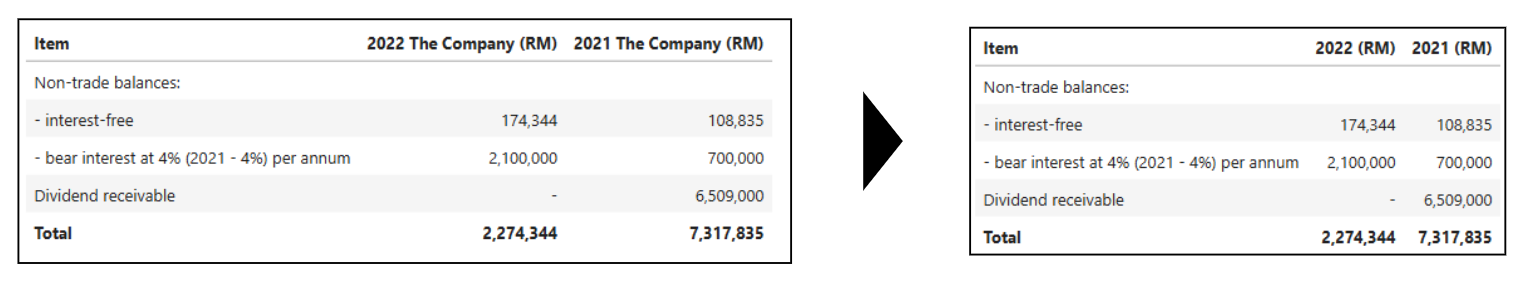}
} \\
\midrule

\RaggedRight
\textbf{Correct Item Order} \newline 
All cell values and row labels must appear in the correct sequence. &
\RaggedRight
\leavevmode\newline
Multiple item values are shifted and misaligned from their correct positions. \\
\multicolumn{2}{c}{
    \includegraphics[width=\linewidth, keepaspectratio]{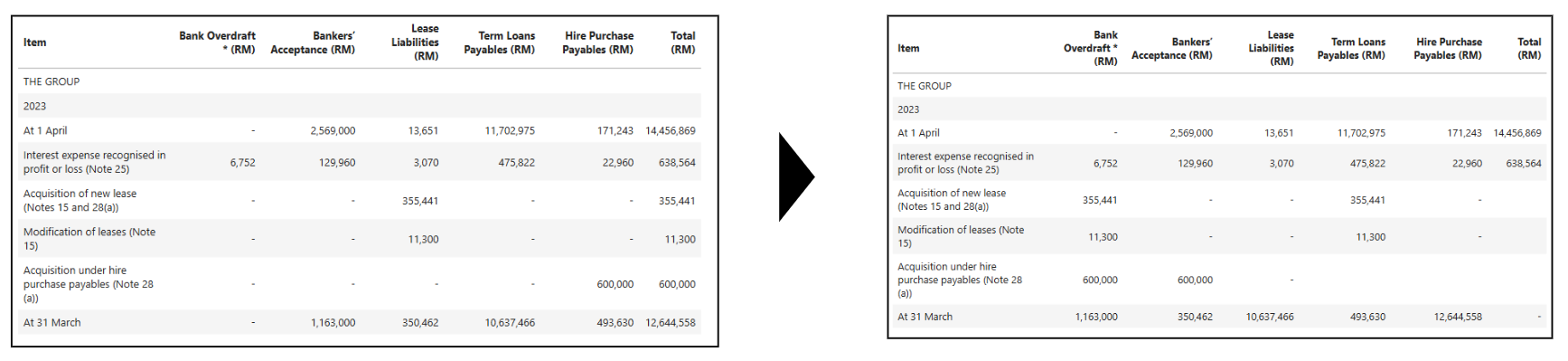}
} \\

\RaggedRight
\textbf{Valid Markdown Formatting} \newline 
Output must adhere to standard Markdown table syntax. &
\RaggedRight
\leavevmode\newline
The model fails to generate correct Markdown syntax for row separators and the header delimiter, causing the table to render as plain text. \\
\multicolumn{2}{c}{
    \includegraphics[width=\linewidth, keepaspectratio]{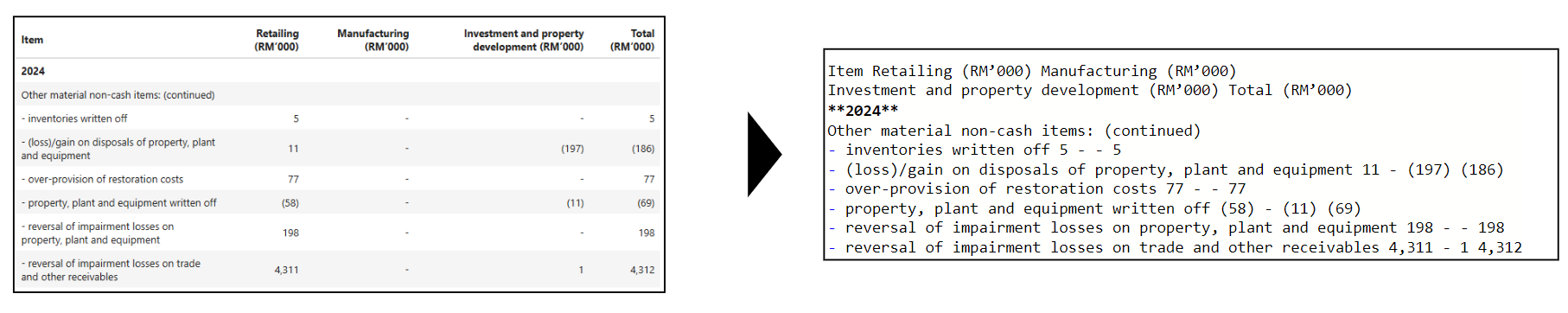}
} \\

\end{longtable}

Recognizing that LLMs may not be infallible, the automated judgments were subsequently subjected to manual verification. This allowed us to correct any occasional inaccuracies or misinterpretations made by the LLM judge, thereby ensuring the final results are highly reliable.

\paragraph{Scoring with Markdown TEDS}
Second, to obtain a single, holistic score for structural and content fidelity, we introduce a custom metric named \textbf{Markdown TEDS}. This metric is built upon the Tree-Edit-Distance-based Similarity (TEDS) framework~\citep{Zhong2019-qm}, which evaluates table accuracy by representing tables as HTML trees and computing the structural edit distance between a prediction and a ground truth. This approach, also used in the evaluation of models such as TableFormer~\citep{Nassar2022-ml}, provides a single score that holistically accounts for both structural and content errors, addressing the limitations of traditional metrics that often overlook major structural issues while over-penalizing minor content mistakes.

Our implementation adapts this powerful tree-based comparison methodology for the nuances of Markdown output. It begins by parsing the raw Markdown and isolating only the table structures, thereby ensuring that any non-tabular text (e.g., headings, paragraphs) does not influence the score. Each extracted table is then converted into a tree representation. To handle typical generation irregularities, our metric incorporates three novel modifications:

\begin{itemize}
    \item \textbf{Table Structure Isolation:} The initial step of our pipeline is a pre-processing stage that programmatically identifies and extracts only the Markdown syntax corresponding to tables. All surrounding text, such as titles or explanatory paragraphs, is disregarded. This ensures that the metric is a pure measure of table structure and content, unaffected by the model's performance on non-tabular generation.

    \item \textbf{Fuzzy Table Merging:} We observed that models sometimes incorrectly fragment a single, large logical table into multiple smaller tables during generation. To handle this, we implemented a merging heuristic. Our system iterates through consecutive tables and merges them if their headers are textually similar above a defined threshold (e.g., >80\% average cell-wise similarity). This allows the metric to correctly reconstruct the intended logical structure from a fragmented prediction before comparison, preventing unfair penalization.

    \item \textbf{Optimal Multi-Table Matching:} A document can contain multiple distinct tables. To evaluate this accurately, our metric calculates a similarity score for every possible pairing of a predicted table with a ground truth table. It then employs the Hungarian algorithm~\citep{Kuhn1955-us} to solve this assignment problem, finding the optimal set of one-to-one matches that maximizes the total similarity. This provides a document-level score based on how closely the predicted tables align with the reference set, with mismatches affecting the overall score.
\end{itemize}

Together, the LLM judge's detailed feedback and the Markdown TEDS score provide a comprehensive evaluation of model performance. The models evaluated included our fine-tuned Qwen2.5-VL-7B, alongside baselines: Qwen2.5-VL-7B (base), Qwen2.5-VL-32B AWQ, Qwen2.5-VL-32B, Qwen2.5-VL-72B AWQ, Keye-VL-8B, GLM-4.1V-9B (Thinking mode), and MiMo-VL-7B-RL, and the proprietary models OpenAI GPT-4o and Gemini 2.5 Flash. The 72B model was evaluated using its AWQ version due to VRAM limitations on our two units of A100 40GB GPU setup; this quantization method has been shown to maintain high fidelity, with a reported performance drop of only \textbf{1.4\%} on the COCO Captioning benchmark ~\citep{Lin2014-xq} for a VLM of comparable size~\citep{Lin2023-ur}. For each of the 100 test cases, we calculated the pass rate for each of the five qualitative criteria and also recorded the Markdown TEDS score. Additionally, we recorded the time taken to generate each Markdown output, measured on a vLLM setup, to evaluate computational efficiency alongside accuracy. Inference times for GPT-4o and Gemini 2.5 Flash were excluded, as inference via API involves different runtime conditions, making direct comparison to local models not directly representative.

\section{Evaluation Results}
\label{sec:results}

We evaluated the performance of our fine-tuned Qwen2.5-VL-7B model against several baseline vision-language models (VLMs), including proprietary models, on our test set of 100 Malaysian financial tables. The evaluation for all self-hosted models was conducted using a vLLM setup hosted on two units of A100 40GB GPUs. We employed a dual-metric approach for a comprehensive assessment: (1) a criteria-based evaluation using OpenAI's \texttt{o3-mini} as an automated judge, and (2) our holistic \textbf{Markdown TEDS} score.

For the criteria-based evaluation, we measured the pass rate (accuracy) for each of the five criteria defined in Section \ref{sec:eval_methodology}. The "Overall Accuracy" is the mean of the five pass rates. The Markdown TEDS score provides a single, unified measure of structural and content fidelity, ranging from 0\% (completely dissimilar) to 100\% (identical).

The results, summarized in Table \ref{tab:eval_results} demonstrate that our fine-tuned Qwen2.5-VL-7B model significantly outperforms all baselines, including proprietary ones, in both evaluation frameworks. It achieves an overall accuracy of 92.20\% and a Markdown TEDS score of 96.53\%.

\begin{table}[H]
\centering
\caption{Performance Comparison of VLM Models on Financial Table Markdownification.}
\label{tab:eval_results}
\renewcommand{\arraystretch}{1.2}
\resizebox{\textwidth}{!}{%
\begin{tabular}{@{} llrrrrrrrr @{}}
\toprule
\textbf{Model} & \textbf{Type} & \textbf{Row (\%)} & \textbf{Col. (\%)} & \textbf{Headers (\%)} & \textbf{Order (\%)} & \textbf{Format (\%)} & \textbf{Overall (\%)} & \textbf{TEDS (\%)} & \textbf{Time (s)} \\
\midrule
\multicolumn{10}{@{}l}{\textit{Open-Source Baselines (Self-Hosted)}} \\
\midrule
Qwen2.5 VL 7B & Standard & 30.00 & 18.00 & 11.00 & 15.00 & 90.00 & 32.80 & 52.08 & 1291.22 \\
MiMo VL 7B RL & Reasoning & 63.00 & 53.00 & 46.00 & 50.00 & 79.00 & 58.20 & 58.56 & 3724.73 \\
Keye VL 8B & Reasoning & 24.00 & 26.00 & 16.00 & 15.00 & 80.00 & 32.20 & 42.06 & 4491.11 \\
GLM 4.1V 9B & Reasoning & 39.00 & 41.00 & 30.00 & 34.00 & 84.00 & 45.60 & 45.97 & 4949.04 \\
Qwen2.5 VL 32B AWQ & Quantized & 52.00 & 61.00 & 45.00 & 42.00 & 98.00 & 59.60 & 66.04 & 2323.39 \\
Qwen2.5 VL 32B & Standard & 58.00 & 74.00 & 59.00 & 53.00 & 99.00 & 68.60 & 71.20 & 2681.51 \\
Qwen2.5 VL 72B AWQ & Quantized & 75.00 & 86.00 & 74.00 & 67.00 & 97.00 & 79.80 & 74.72 & 2974.71 \\
\midrule
\multicolumn{10}{@{}l}{\textit{Proprietary Baselines (API-based)}} \\
\midrule
OpenAI GPT-4o & Proprietary & 61.00 & 62.00 & 54.00 & 54.00 & 95.00 & 65.20 & 74.41 & - \\
Gemini 2.5 Flash & Proprietary & 81.00 & 90.00 & 72.00 & 79.00 & 90.00 & 82.40 & 79.19 & - \\
\midrule
\rowcolor{boxbg} \textbf{Qwen2.5 VL 7B (Ours)} & \textbf{Finetuned} & \textbf{94.00} & \textbf{93.00} & \textbf{84.00} & \textbf{90.00} & \textbf{100.00} & \textbf{92.20} & \textbf{96.53} & \textbf{804.67} \\
\bottomrule
\end{tabular}%
}
\end{table}

The effectiveness of our domain-specific fine-tuning is demonstrated when comparing our model to its base counterpart, Qwen2.5-VL-7B. Our model achieves an overall accuracy of 92.20\% and a TEDS score of 96.53\%, representing a substantial improvement over the base model’s 32.80\% accuracy and 52.08\% TEDS.

Our fine-tuned model’s superiority is also evident when evaluated against other models in the 7–9B parameter class. It surpasses all reasoning-enabled models in this range, including MiMo-VL-7B-RL (58.20\% Overall, 58.56\% TEDS), GLM-4.1V-9B (45.60\% Overall, 45.97\% TEDS), and Keye-VL-8B (32.20\% Overall, 42.06\% TEDS) indicating that domain-specific adaptation is more effective for this task than generalized multimodal reasoning.

In addition, the model outperforms significantly larger open-source alternatives. Qwen2.5-VL-32B achieves 68.60\% Overall and 71.20\% TEDS, while Qwen2.5-VL-72B AWQ reaches 79.80\% Overall and 74.72\% TEDS, both of which are well below our 7B model’s performance. This highlights the efficiency of targeted fine-tuning in achieving high performance without reliance on larger model scales.

Our model also surpasses widely used proprietary models. It achieves higher accuracy than both Gemini 2.5 Flash (82.40\% Overall, 79.19\% TEDS) and OpenAI’s GPT-4o (65.20\% Overall, 74.41\% TEDS). This demonstrates that a lightweight, specialized open-source model can deliver state-of-the-art results for this task, offering a more efficient and transparent alternative to generalist, closed-source models.

In terms of processing time, our model is the fastest among all self-hosted models, completing the full test set in 804.67 seconds. By contrast, the base Qwen2.5-VL-7B required 1291.22 seconds and often generated redundant or irrelevant output until reaching the token limit. The fine-tuned model consistently produced accurate Markdown, avoiding unnecessary generation. Reasoning-heavy models such as GLM-4.1V-9B (4949.04s), Keye-VL-8B (4491.11s), and MiMo-VL-7B-RL (3724.73s) exhibited significantly longer runtimes due to their reasoning overhead.

These results underscore the value of domain-specific fine-tuning for specialized tasks such as financial document understanding. Our approach delivers a lightweight, high-performance model that not only achieves state-of-the-art accuracy but also outpaces significantly larger and more complex models in both effectiveness and efficiency.

\section{Discussion}
\label{sec:discussion}

The fine-tuned Qwen2.5-VL-7B model demonstrates a strong capability in converting Malaysian financial tables into Markdown format, handling diverse layouts, rotated orientations, hierarchical headers, and implicit structures with high precision. Our dual-metric evaluation framework confirms its superior performance. On the criteria-based assessment, it achieved an overall accuracy of 92.20\%, with standout results in row alignment (94.00\%) and Markdown compliance (100.00\%). This is further supported by its \textbf{Markdown TEDS score of 96.53\%}, indicating that its outputs are not only discretely correct but also holistically and structurally almost identical to the ground truth. These results stem from a carefully constructed 2,152-sample training dataset, which includes 30\% rotated examples, and a focused LoRA fine-tuning strategy performed on 2$\times$ A100 40GB GPUs.

Relative to the Qwen2.5-VL-7B base model (32.80\% accuracy, 52.08\% TEDS), the fine-tuned version exhibits a dramatic improvement in both accuracy and structural fidelity. It also shortens inference time from 1291.22 seconds to 804.67 seconds by avoiding the redundant token generation that plagues the base model. Crucially, our model outperforms significantly larger architectures, including the Qwen2.5-VL-32B (68.60\% accuracy, 71.20\% TEDS) and the Qwen2.5-VL-72B AWQ (79.80\% accuracy, 74.72\% TEDS).

Furthermore, the model surpasses all evaluated reasoning-focused VLMs and even proprietary models. It scored significantly higher than both Gemini 2.5 Flash (82.40\% accuracy, 79.19\% TEDS) and OpenAI's GPT-4o (65.20\% accuracy, 74.41\% TEDS), demonstrating that for this domain-specific task, a specialized, lightweight model can be more effective than massive, generalist models. This result provides strong evidence that for specialized structural understanding, targeted training is a more effective and efficient strategy than relying on general-purpose, reasoning-heavy, or closed-source approaches. The fine-tuning process, supported by a stable training configuration, presents a scalable and efficient solution for financial document parsing, reducing dependence on commercial APIs.

\section{Limitations and Future Work}
\label{sec:limitations_future_work}

While our fine-tuned model demonstrates state-of-the-art performance, this study has several limitations that present opportunities for future work. The training process was conducted on two A100 40GB GPUs, which imposed computational constraints on our experimental design. This limited our ability to explore larger batch sizes, which could potentially improve training stability and final model performance, or to conduct more extensive hyperparameter sweeps. Due to these resource constraints, we adopted a parameter-efficient fine-tuning method (LoRA). While effective under limited hardware, LoRA restricts the extent of model adaptation compared to full fine-tuning.

Additionally, while our 2,152-sample dataset proved effective, a larger and more varied dataset could further enhance the model's robustness and ability to handle rare edge cases. Similarly, our 100-sample test set, while carefully curated to represent diverse and complex cases, is relatively small. A more extensive evaluation benchmark would be required to draw broader conclusions about the model's performance across the entire spectrum of financial reporting.

Building on this work, several avenues for future research present exciting opportunities. The dataset could be expanded to encompass a wider variety of table formats found in documents beyond financial reports, such as academic papers, technical documentation, or regulatory filings, to improve the model's generalizability. Furthermore, exploring direct generation of structured HTML represents a promising avenue for enhancing the model's capabilities. While Markdown provides a highly readable and effective representation for the majority of tables, HTML offers native support for the most complex structures via \texttt{rowspan} and \texttt{colspan} attributes. The utility of this rich format for explicitly encoding merged cells is demonstrated by its adoption in prominent table recognition benchmarks, such as PubTabNet~\citep{Zhong2019-qm}, which leverages HTML to capture intricate scientific table layouts. Finally, integrating this high-fidelity table extraction model into downstream analytical pipelines, such as financial question-answering or summarization systems, would be a valuable next step in creating end-to-end document intelligence solutions.

\section{Conclusion}
\label{sec:conclusion}

This work introduces a practical solution for converting tables from Malaysian audited financial reports into Markdown by fine-tuning an open-source vision-language model (VLM). By targeting the structural challenges found in these documents, such as rotated layouts, multi-level headers, and implicit structures, we demonstrate that standard, off-the-shelf VLMs, including those with general reasoning capabilities and even proprietary SOTA models, are insufficient for reliable table understanding in this specialized domain without adaptation.

Our fine-tuned Qwen2.5-VL-7B model, trained using a domain-specific dataset of 2,152 image-text pairs and optimized with LoRA, achieves state-of-the-art performance. Validated through a dual-metric framework, it attains a \textbf{92.20\% overall accuracy} on a criteria-based LLM-as-a-judge evaluation and a \textbf{96.53\% Markdown TEDS score}, confirming its exceptional structural and content fidelity. This performance substantially surpasses its Qwen2.5-VL-7B base model and larger alternatives such as the Qwen2.5-VL-72B AWQ, while also being significantly faster than these self-hosted baselines. Notably, its accuracy also exceeds that of prominent proprietary models such as GPT-4o and Gemini 2.5 Flash.

This outcome underscores the immense value of targeted fine-tuning on well-structured datasets for specialized document understanding tasks. Our "Markdownification" pipeline provides a practical and interpretable intermediate representation that facilitates downstream analysis. Ultimately, this work highlights that high accuracy in financial table extraction does not require massive proprietary models or complex reasoning mechanisms. With strategic fine-tuning, open-source VLMs can match or exceed state-of-the-art performance, offering a transparent, cost-effective, and auditable alternative for financial document processing in real-world applications.

\newcommand{\huggingfacelogo}{%
    \raisebox{-0.7ex}{\includegraphics[height=3ex]{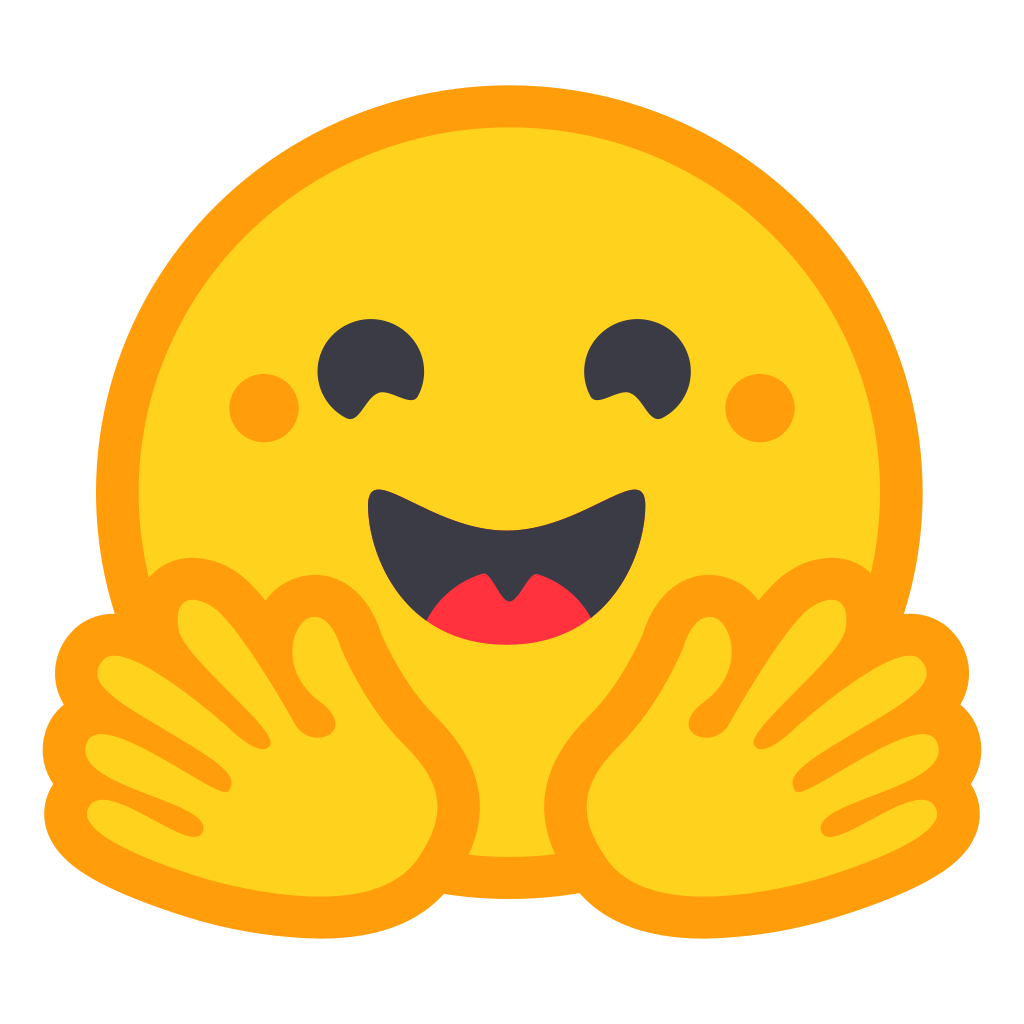}}%
    ~%
}

\section*{Data Availability}
\label{sec:data_availability}

The full dataset used for the experiments in this paper consists of 2,152 samples. To promote transparency and enable further research, we have publicly released two key datasets: a development sample and our full evaluation benchmark.

\begin{itemize}
    \item \textbf{Training and Development Subset:} A representative sample of our development data, derived from 100 companies. It contains 699 total entries, comprising 538 base tables and 161 rotated augmentations (a 30\% augmentation rate). This is intended for users who wish to explore the data or replicate our fine-tuning process on a smaller scale. 
    It is available on \href{https://huggingface.co/datasets/jinkhye/MyFinMarkdown-sample}{\huggingfacelogo MyFinMarkdown-sample}.

    \item \textbf{Evaluation Benchmark:} The complete 100-sample test set used to generate the final results reported in Section \ref{sec:results}. This dataset can be used to benchmark other models directly against our findings.
    It is available on \href{https://huggingface.co/datasets/jinkhye/MyFinMarkdown-bench}{\huggingfacelogo MyFinMarkdown-bench}.
\end{itemize}

\newcommand{\githublogo}{\raisebox{-0.1ex}{\color{primaryblue}\faGithub}~}
\section*{Code Availability}
\label{sec:code_availability}

The scripts for our LLM-as-a-judge evaluation and the custom Markdown Tree-Edit-Distance-based Similarity (TEDS) metric are publicly available. An example implementation can be found in our GitHub repository. \href{https://github.com/jinkhye/MyFinMarkdown}{\githublogo MyFinMarkdown}

\clearpage

\section*{Acknowledgement}

I would like to express my sincere gratitude to \textbf{Huawei}, and in particular the \textbf{APAC Regional Artificial Intelligence Center of Excellence (AICOE) team}, for the opportunity to undertake this internship and contribute to a meaningful project. I also wish to thank the \textbf{Faculty of Computer Science and Information Technology, Universiti Malaya}, for fostering the academic environment that supported this work.

I am especially thankful for the mentorship and guidance provided throughout this journey. My gratitude extends to my mentor, Mr. Kang Chin Shen for his valuable support during the initial stages of my internship, and to my university supervisor, Ts. Dr. Riyaz Ahamed, for his unwavering support and academic oversight. Their collective guidance has been instrumental to the development and completion of this research.

\clearpage

\appendix
\section{Appendix}
\label{sec:appendix}
\addcontentsline{toc}{section}{Appendix A}

This appendix provides the prompts used for both generating the initial Markdown output from Malaysian audited financial reports and for evaluating the quality of the generated Markdown against a ground truth. An example of the ShareGPT data format used for training is also included.

\subsection*{Prompt for Markdown Generation (VLM Fine-tuning)}
\addcontentsline{toc}{subsection}{Prompt for Markdown Generation (VLM Fine-tuning)}

The following system and user prompts were used to instruct the Vision-Language Model (VLM) in generating Markdown from financial table images. This prompt defines the rules for table detection, parsing, header canonicalization, row rendering, and specific handling of section headers within tables, subtotals, and totals.

\begin{tcolorbox}[
    title={System Prompt for Markdownify},
    colback=boxbg,
    colframe=primaryblue!70!white,
    boxrule=0.8pt,
    fonttitle=\bfseries\sffamily\small,
    sharp corners,
    breakable,
    left skip=0pt, right skip=0pt,
    width=\textwidth
]
\begin{verbatim}
You are a PDF-to-Markdown converter specialized in financial tables.
Input: raw text extracted from a image, including titles, section headers, 
footnotes, and any number of tables with varying column layouts.
Output: ONLY clean Markdown.
\end{verbatim}
\end{tcolorbox}

\begin{tcolorbox}[
    title={User Prompt for Markdownify},
    colback=boxbg,
    colframe=primaryblue!70!white,
    boxrule=0.8pt,
    fonttitle=\bfseries\sffamily\small,
    sharp corners,
    breakable,
    left skip=0pt, right skip=0pt,
    width=\textwidth
]
\begin{verbatim}
1. Detect and emit table titles & headers
- Output any standalone title and headers if any preceding the section 
containing the table  → Markdown header (#).

2. Parse each table
a. Identify the header row (first row containing multiple columns).
b. Extract:
- All year tokens (YYYY).
- Any entity tokens (Group, Company). 
- Any measure tokens (Number of shares, Amount, currency codes).
- A “Note” column if present, and an “Item” or description column.
c. Only when “Group”/“Company” information is presented in the header, 
include the Group/Company in the parsed header such as: "2023 Group
(CUR'000)".
- For non-“Group/Company” cases, use back the original headers in the 
source table.
d. Never transpose the table.

3. Build a canonical header
- Always include a header for all columns.
- Analyse and replace the header with a suitable title ONLY IF the header 
is missing. 
- Include a Note column after the first description column only if the 
source table has one.

4. Render rows
- Detect columns visually — based on vertical alignment, grid lines, or 
spacing.
- Split each row cell by column position. Maintain alignment 
between headers and data cells—each row must have the exact same number 
of cells as the header.
- If a row is shorter than expected, fill missing cells with '-'. Detect 
columns visually — based on vertical alignment, grid lines, or spacing.
- Preserve note numbers in the Note column.
- Negative numbers keep their minus sign or parentheses, but remain 
in-cell.

5. Section headers inside tables
- If a row has a single merged label (e.g. “CASH FLOWS FROM FINANCING 
ACTIVITIES”), render as:
| CASH FLOWS FROM FINANCING ACTIVITIES |||||

6. Subtotals & Totals
- If a row represents “Total” or “Net”, bold the entire row label and 
all its numeric values.

7. No explanations
- Output only the final Markdown containing tables and text.
- Attempt to follow the structure of the source tables as closely as 
possible.
- Do not include any commentary, analysis, or non-Markdown text.


<Example1 - missing header on first column>
Source headers:
| Group |
| 2024 CUR'000 |2023 CUR'000 |

Canonical header:
| <Header>  | 2024 Group (CUR'000) | 2023 Group (CUR'000) |
*Replace <Header> with a suitable name if missing
</Example1>

<Example2>
Source headers:
| The Group | | The Company | |
| At cost | Note | 2024 CUR'000 |2023 CUR'000 | 2024 CUR'000 | 
2023 CUR'000 |

Canonical header:
| At cost | Note | 2024 The Group (CUR'000) | 2023 The Group (CUR'000) | 
2024 The Company (CUR'000) | 2023 The Company (CUR'000) |
</Example2>

Return the markdown representation of this document as if you were 
reading it naturally.
\end{verbatim}
\end{tcolorbox}

\subsection*{ShareGPT Data Format Example}
\addcontentsline{toc}{subsection}{ShareGPT Data Format Example}
The training dataset is structured in the ShareGPT format. For our VLM fine-tuning, each conversation consists of a single turn with a system prompt, a user prompt (containing the raw text and an image placeholder), and an assistant's response (the ground truth Markdown). The following is a representative example from our dataset, truncated for brevity.

\begin{tcolorbox}[
    title={ShareGPT Data Format Example},
    colback=boxbg,
    colframe=primaryblue!70!white,
    boxrule=0.8pt,
    fonttitle=\bfseries\sffamily\small,
    sharp corners,
    breakable,
    left skip=0pt, right skip=0pt,
    width=\textwidth
]
\footnotesize
\begin{verbatim}
[
    {
        "role": "system",
        "content": "You are a PDF-to-Markdown converter..."
    },
    {
        "role": "user",
        "content": "1. Detect and emit table titles & headers...\n\n
        Return the markdown representation...\n\n
        ## Following is the raw text from the image:\n\n
        INTEGRATED REPORT 2023\r\n115SECTION 9: FINANCIAL STATEMENT\r\n...\n\n
        <image>"
    },
    {
        "role": "assistant",
        "content": "## STATEMENTS OF FINANCIAL POSITION\r\n
        ### AS AT 31 DECEMBER 2023\r\n\r\n
        | Item | Note | ... | 2022 Company (RM’000) |\r\n
        |:---|:---|---:|---:|---:|---:|\r\n
        ...\r\n
        | **TOTAL EQUITY** | - | ... | **693,612** |\r\n\r\n
        The notes on pages 121 to 202..."
    }
]
\end{verbatim}
\end{tcolorbox}

\subsection*{Prompt for LLM-based Evaluation (LLM Judge)}
\addcontentsline{toc}{subsection}{Prompt for LLM-based Evaluation (LLM Judge)}

The following defines the prompt template used for the LLM-based evaluation process. This prompt guides the LLM (OpenAI o3-mini) to assess the accuracy and formatting of the generated Markdown against a ground truth based on five specific criteria.

\begin{tcolorbox}[
    title={LLM Judge Prompt Template},
    colback=boxbg,
    colframe=primaryblue!70!white,
    boxrule=0.8pt,
    fonttitle=\bfseries\sffamily\small,
    sharp corners,
    breakable,
    left skip=0pt, right skip=0pt,
    width=\textwidth
]
\begin{verbatim}
You are an expert Markdown evaluator for financial statement pages. Assess 
the 'Actual Output' against the 'Expected Output' based on the rules below.

Return:
- "criteria": A dictionary with keys for each criterion and boolean values 
(True if met, False if not).

---

**Evaluation Criteria for Markdown Documents:**

1. **Correct Row Count**: All tables in the Actual Output have the same 
number of rows as in the Expected Output.
2. **Correct Column Count**: All tables in the Actual Output have the same 
number of columns as in the Expected Output.
3. **Semantically Accurate Headers**: All table headers in the
Actual Output convey the same meaning as those in the Expected Output 
(minor wording differences are acceptable if the intent is preserved).
4. **Correct Item Order**: All table items and cell values in the Actual 
Output maintain the same order as in the Expected Output without shifts or 
misplacements.
5. **Valid Markdown Formatting**: The Actual Output uses correct Markdown 
syntax (e.g., proper table structure, header syntax) consistent with the 
Expected Output.

---

Test Case:
Actual Output:
{predicted}

Expected Output:
{gold}

---
**Example Response (LLM must follow this strict format):**
{{
    "criteria": {{
        "Correct Row Count": false,
        "Correct Column Count": true,
        "Semantically Accurate Headers": false,
        "Correct Item Order": true,
        "Valid Markdown Formatting": true,
    }}
}}
\end{verbatim}
\end{tcolorbox}

\clearpage

\bibliographystyle{unsrtnat}  
\bibliography{ref-extracts}  

\end{document}